\documentclass[%
reprint,
 amsmath,amssymb,
 aps,
]{revtex4-2}

\usepackage{graphicx}
\usepackage{dcolumn}
\usepackage{bm}
\usepackage{color}

\usepackage{subfigure}

\begin{document}

\preprint{APS/123-QED}

\title{Constraining the stochastic gravitational wave background using the future lunar seismometers}

\author{Han Yan$^{1,2}$}
\author{Xian Chen$^{1,2}$}
  \email{Corresponding author. \\ xian.chen@pku.edu.cn}
\author{Jinhai Zhang$^{3}$} 
\author{Fan Zhang$^{4,5}$}
\author{Lijing Shao$^{2}$}
\author{Mengyao Wang$^{5}$}

\affiliation{$^{1}$Department of Astronomy, School of Physics, Peking University, 100871 Beijing, China}

\affiliation{$^{2}$Kavli Institute for Astronomy and Astrophysics at Peking University, 100871 Beijing, China}

\affiliation{$^{3}$Institute of Geology and Geophysics, Chinese Academy of Sciences, Beijing 100029, China}

\affiliation{$^{4}$Institute for Frontiers in Astronomy and Astrophysics, Beijing Normal University, Beijing 102206, China}

\affiliation{$^{5}$Department of Astronomy, Beijing Normal University, Beijing 100875, China}

\date{\today}

\begin{abstract}
Motivated by the old idea of using the moon as a resonant gravitational-wave (GW) detector, as well as the recent updates 
in modeling the lunar response to GWs, 
we re-evaluate the feasibility of using a network of lunar seismometers 
to constrain the stochastic GW background (SGWB). 
In particular, using the updated model of the lunar response, 
we derive the pattern functions for the two polarizations of GW. 
With these pattern functions, we 
further calculate the overlap reduction functions for a network of lunar seismometers, 
where we have relaxed the conventional assumption that lunar seismometers 
are perfectly leveled to measure only the vertical acceleration.
We apply our calculation to two future lunar projects, namely, Chang'e and
the Lunar Gravitational-Wave Antenna (LGWA). 
We find that the two projects could constrain the SGWB
to a level of $\Omega_{\text{GW}}^{\text{Chang'e}} <  2.4 \times 10^{2}$ and
$\Omega_{\text{GW}}^{\text{LGWA}} <  2.0 \times 10^{-10}$, respectively,
which corresponds to a signal-to-noise ratio of SNR $=3$.
These results are better than the
constraints placed previously 
	on the SGWB in the mid-frequency band (around $10^{-3}- 10~\text{Hz}$)
by various types of experiments.
\end{abstract}

\maketitle


\section{\label{sec:intro}Introduction }

The idea of using the earth or the moon as a resonant gravitational-wave (GW) detector is an
old one \cite{1960PhRv..117..306W,1968PhT....21d..34W,1974ApJ...187L..49M}. Early estimations suggest that the moon might be sensitive to GW at the frequencies higher than 1 milli-Hertz (mHz) \cite{1973grav.book.....M}, due to its fundamental quadrupole mode. Meanwhile, detailed theory about the seismic response to GW gradually developed from 1960s to 1980s \cite{1968PhT....21d..34W,1983NCimC...6...49B}.
The sensitivity in such a
mid-frequency band (around $10^{-3}- 10~\text{Hz}$) is a good complement to the current ground-based GW
detectors like LIGO/Virgo/KAGRA
\cite{1992Sci...256..325A,2015CQGra..32b4001A,2019NatAs...3...35K}, which are
tuned to detect the GWs in the audible band ($10-10^3$ Hz). It is also
complementary to the future space-borne interferometers such as LISA
\cite{2024arXiv240207571C}, TianQin \cite{2016CQGra..33c5010L} and Taiji \cite{2021PTEP.2021eA108L}, which are
sensitive to mHz GWs.
 
Among the various signals in the mid-frequency band \cite{2020CQGra..37u5011A},
stochastic GW background (SGWB) is a persistent one \cite{1999PhRvD..59j2001A}.
Such a background could form during the early cosmic inflation
\cite{1988PhRvD..37.2078A,2006JCAP...04..010E,2018CQGra..35p3001C} or due to
the evolution of a population of binaries in the Milky Way
\cite{2003MNRAS.346.1197F,2011PhRvD..84h4004R,2021PhRvD.104d3019K}. In the most optimistic scenarios, the predicted energy density from these sources amounts to $\Omega_{\text{GW}} \simeq
10^{-8} \sim  10^{-10}$ in the mid-frequency region. 

Several earlier works tried to constrain the SGWB in the mid-frequency band by
analyzing the data from the seismometer networks on the earth or the moon.
They studied both the free-surface effect  \cite{2014PhRvL.112j1102C} (hereafter
CH14a) and the normal modes \cite{2014PhRvD..90d2005C} (hereafter CH14b) of the
earth, as well as the normal modes of the moon \cite{2014PhRvD..90j2001C}
(hereafter CH14c). However, limited by the sensitivities of the seismometers,
the results are insufficient to put stringent constraint on the SGWB, even
though they are already better than (or at least comparable to) the constraints placed by 
other types of experiments, such as ULYSSES \cite{1995A&A...296...13B}, LP \cite{2001Icar..150....1K}, GPS \cite{2013qopu.confE..29A},
GRACE-FO \cite{2020PhRvD.101j2004R}, and Tianwen-I \cite{2024arXiv240206096B}.  Therefore, different designs of lunar
seismometers that are underway can boost the sensitivity in this band, such as the
Lunar Gravitational-Wave Antenna (LGWA)
\cite{2021ApJ...910....1H} and other designs (e.g., \cite{2023SCPMA..6609513L}).

Three recent progresses motivated us to revisit the calculation of the
sensitivity of lunar seismometers to the SGWB. First, 
the physical interpretation and theoretical formulation of the lunar response to
GW have evolved in recent years
\cite{2019PhRvD.100d4048M,PhysRevD.109.064092,2024arXiv240316550B}, and the
response functions of the moon have been updated
\cite{PhysRevD.109.064092,2023arXiv231211665K,2024arXiv240305118B}.
Conventionally, the lunar response to GW has been understood in two different
ways, either from the tidal acceleration or from the local shear force induced
by GW (the latter one corresponds to the result of an effective-field-theory
approach \cite{2024arXiv240316550B}).  In light of the recent studies, it becomes clear that the
latter approach results in a response function which is directly proportional
to the readout of a lunar seismometer.  Second, deployment of new seismometers
on the moon is imminent, including China's Chang'e 7 and 8 projects
\cite{2019Sci...365..238L,2020LPI....51.1755Z,10.1093/nsr/nwad329} which are scheduled in 2026 and
2028, and NASA's Farside Seismic Suite (FSS) scheduled in 2026 \cite{2024LPICo3040.2354P}. The India's Chandrayan-3 project launched in 2023 also had the Instrument for Lunar Seismic Activity (ILSA) \cite{2024P&SS..24205864K}. Understanding their sensitivity to the SGWB is urgent.
Third, depending on the design of seismometer and the deployment mechanism, 
future lunar seismometers may not be perfectly leveled, and one seismometer may be able to
measure the accelerations in multiple directions, but the previous works 
commonly assumed that the seismometer arrays are responding only to the vertical acceleration. A related matter is that seismometers with different orientations
are sensitive to different polarization states of the incoming GWs. But the
previous  works are mostly based on a single polarization state. Here, we aim
to address the above issues.

The paper is organized as follows.  In Section~\ref{sec:theo}, we review the
method of calculating the lunar response to GW, and the theory of detecting the
SGWB using multiple detectors. In Section~\ref{sec:ORF}, we
calculate the overlap reduction function (ORF) of lunar-seismometer array,
paying special attention to the configuration in which seismometers point to
arbitrary directions. In Section~\ref{sec:appl}, we apply our theoretical
framework to two future projects, namely, Chang'e and LGWA, to evaluate their
abilities in constraining the SGWB.  Finally, in Section~\ref{sec:con}, we
summarize our results and discuss possible aspects for future research.
Throughout the paper we will adopt the International System of Units and the
Minkowski metric $\eta _{\mu \nu } = diag \left ( -1,1,1,1 \right ) $, unless
mentioned otherwise. 

\section{\label{sec:theo}Basic Theory}

In this section, we first review the surface response solution of a radially
heterogeneous elastic sphere (which is a good first-step approximation to the
real case of the moon) to linearly-polarized GWs, basically following
\cite{2019PhRvD.100d4048M} (hereafter Ma19) and \cite{PhysRevD.109.064092}
(hereafter Yan24). We also generalize this solution to account for GWs with
arbitrary polarization states. Based on this solution we derive two pattern
functions corresponding to the two polarizations of GW.  Next, we briefly
describe the calculation of the signal-to-noise ratio (SNR) of SGWB for an
array of GW detectors, which introduces an ORF. Using
this SNR, we evaluate the capability of using lunar seismometers to constrain
the SGWB.

\subsection{\label{sec:resptheo}Surface response of the moon}

As we have clarified in Yan24, given a linearly polarized GW,  
\begin{eqnarray}
    \mathbf{h} && =  \Re \left \{ h_{0} \epsilon _{ij} e^{i\left ( \omega _{g}t - \vec{k}_{g}\cdot \vec{r}     \right ) }    \right \} \nonumber \\ \vec{k}_{g} &&=\left ( 0,0,\omega _{g}/c\right ) \nonumber \\ \epsilon _{ij} && = \begin{bmatrix}
 1 & 1 & 0\\
 1 & -1 & 0\\
 0 & 0 & 0
	\end{bmatrix} ~, \label{GWtensor}
\end{eqnarray}
the surface response of a radially heterogeneous elastic sphere in the long-wavelength approximation can be written as: 
\begin{eqnarray}
    \vec{\xi } \left ( \theta ,\varphi ,t \right )  = && h_{0} \cos \left ( \omega _{g}t  \right ) \times \nonumber\\ &&\Bigg[ T_{r} \sum_{m}  f^{m}   \mathcal{Y}_{2m} \left ( \theta ,\varphi  \right ) \hat{e}_{r} \nonumber \\ && + T_{h} \sum_{m}  f^{m} \partial _{\theta }  \mathcal{Y}_{2m} \left ( \theta ,\varphi  \right ) \hat{e}_{\theta }  \nonumber \\
    &&  +  T_{h} \sum_{m}  f^{m}  \frac{\partial _{\varphi  }  \mathcal{Y}_{2m} \left ( \theta ,\varphi  \right )}{\sin \theta } \hat{e}_{\varphi   } \Bigg] ~,\label{eq:xi}
\end{eqnarray}
where we are evaluating the response at the location of $R\left ( \sin \theta \cos \varphi ,\sin \theta \sin \varphi, \cos \theta  \right )$ ($R$ is the lunar radius).
Following the definition in Yan24, $T_{r}$ and $T_{h}$ are the radial and
horizontal response functions at the GW frequency $\omega _{g}/2\pi$. They are independent
of the location of a detector or the propagation direction of GW.
$\mathcal{Y}_{2m}$ is the real spherical harmonics, and $f^{m}$ is a function
depending on the direction of the propagation of GW:
\begin{eqnarray}
    f^{m} =  4\sqrt{\frac{\pi }{15} } \times \left ( \delta_{m, 2}+  \delta_{m,-2} \right ) ~.  \nonumber
\end{eqnarray}
The definitions of three
base vectors in spherical coordinates are respectively
\begin{eqnarray}
    \hat{e}_{r} &&= \left ( \sin \theta \cos \varphi ,\sin \theta \sin \varphi, \cos \theta  \right ) ~,\nonumber \\ \hat{e}_{\theta } &&= \left ( \cos \theta \cos \varphi ,\cos \theta \sin \varphi, -\sin \theta  \right )  ~,\nonumber \\ \hat{e}_{\varphi  } &&= \left ( - \sin \varphi , \cos \varphi, 0  \right ) ~.
\end{eqnarray}
Notice that the tensor $\mathbf{h}$ used above refers to one single polarization
state with a single frequency, which can also be represented by the commonly used polarization states $h_{+}$ and $h_{\times}$ as $h_{+} + h_{\times}$. Another polarization state $h_{+} - h_{\times}$
is discussed in the Appendix. Therefore, the displacement solution,
Eq.~(\ref{eq:xi}), although widely used in the past, applies only to a single
polarization. 

The displacement solution can be written into a more concise form,
\begin{eqnarray}
    \vec{\xi } \left ( \theta , \varphi, t  \right )  = 2T_{h} \mathbf{h}   \cdot \hat{e}_{r} + \left ( T_{r}-2T_{h}   \right ) \left ( \hat{e}_{r} \cdot \mathbf{h}   \cdot \hat{e}_{r} \right )\hat{e}_{r} ~, \label{eq:resp}
\end{eqnarray}
by combining Eqs.~(22) and (25) in Yan24.
The factor $2$ before $T_{h}$ results from the behavior of the spherical
harmonics (see Yan24 for more discussions). 
The advantage of using the latter form is that it unifies the solutions to
different polarizations. The proof is given in the Appendix.
Thus, when considering a more general GW tensor:
\begin{eqnarray}
    \mathbf{h} &&= \left ( h_{+} \mathbf{e}_{+} + h_{\times } \mathbf{e}_{\times } \right ) e^{i\omega_{g}  t} \nonumber \\
    \mathbf{e}_{+} && = \begin{bmatrix}
 1 & 0 & 0\\
 0 & -1 & 0\\
 0 & 0 & 0
	\end{bmatrix} \nonumber \\
\mathbf{e}_{\times } && = \begin{bmatrix}
 0 & 1 & 0\\
 1 & 0 & 0\\
 0 & 0 & 0
	\end{bmatrix}~,
\end{eqnarray}
we can write the surface response detected by an accelerometer along the direction $\hat{e}_{\rm det }$ as
\begin{eqnarray}
    \xi = \vec{\xi } \left ( \theta , \varphi, t  \right )  \cdot \hat{e}_{\rm det }=\left (F_{+} h_{+} +  F_{\times } h_{\times}\right ) e^{i\omega_{g}  t} ~, 
\end{eqnarray}
in which the two pattern functions are
\begin{eqnarray}
    F_{+} =&& 2T_{h} \left ( \hat{e}_{\rm det } \cdot \mathbf{e}_{+}    \cdot \hat{e}_{r} \right ) \nonumber\\&& + \left ( T_{r}-2T_{h}   \right ) \left ( \hat{e}_{r} \cdot \mathbf{e }_{+}   \cdot \hat{e}_{r} \right )\hat{e}_{r} \cdot \hat{e}_{\rm det } \nonumber \\ F_{\times} = && 2T_{h} \left ( \hat{e}_{\rm det } \cdot \mathbf{e}_{\times}    \cdot \hat{e}_{r} \right ) \nonumber\\&& + \left ( T_{r}-2T_{h}   \right ) \left ( \hat{e}_{r} \cdot \mathbf{e }_{\times}   \cdot \hat{e}_{r} \right )\hat{e}_{r} \cdot \hat{e}_{\rm det } ~. \label{eq:patt}
\end{eqnarray}

For example, if all accelerometers measure the vertical acceleration, we can set
$\hat{e}_{\rm det} = \hat{e}_{r }$ and the pattern functions 
become
\begin{eqnarray}
    F^{\text{CH14c}}_{A} = T_{r} \left ( \hat{e}_{r} \cdot \mathbf{e }_{A}   \cdot \hat{e}_{r} \right ),\quad A = + , \times ~. \label{eq:pattJH}
\end{eqnarray}
These functions recover the results given in CH14c.

\subsection{Theory of detecting SGWB}

For two detectors that have uncorrelated noise but have the same noise spectral density $S_{n} \left ( f \right ) $, given a SGWB with a dimensionless energy spectral density $ \Omega _{\text{GW}} \left ( f \right )  $, the SNR in the mid-frequency region is \cite{2009GReGr..41.1667H}:
\begin{eqnarray}
    \text{SNR} = \frac{3H_{0}^{2} }{4\pi ^{2} } \left [ 2T\int_{10^{-3}~\rm Hz}^{10~\rm Hz}  \mathrm{d}f \Gamma ^{2} \left ( f \right ) \frac{\Omega _{\text{GW}}^{2}  \left ( f \right )}{f^{6}S_{n}^{2}\left ( f \right )   }   \right ] ^{1/2} ~, \label{eq:snr}
\end{eqnarray}
where $T$ is the observational period, and $H_{0}$ is the Hubble parameter. 
In the above equation,
the ORF $\Gamma \left ( f \right )$ for two detectors (marked with subscript 1 and 2) is defined as:
\begin{eqnarray}
    \Gamma \left ( f \right ) =&& \int \frac{\mathrm{d}^{2}\hat{n}  }{4\pi } \int \frac{\mathrm{d} \psi }{2\pi } \nonumber \\ &&\times\left [ \sum_{A}F_{1}^{A}\left ( \hat{n}  \right ) F_{2}^{A}\left ( \hat{n}  \right ) \text{exp}\left \{ i2\pi f \hat{n}\cdot \frac{\Delta \vec{x}  }{c} \right \}    \right ]  , \label{eq:ORF}
\end{eqnarray}
where $\hat{n}$ is an unit vector along the GW wave vector, $\psi$ is the
polarization angle, $A = +, \times$ represent different polarization states,
and $\Delta \vec{x}$ is the separation between two detectors. Notice that for
the lunar seismometer array in the frequency range lower than $10 \text{Hz}$,
the exponential term in Eq.~(\ref{eq:ORF}) can be ignored. This is
partly because of the long wavelength, and partly of the angular symmetry of
the wave vector $\hat{n}$ during the integration and averaging \footnote{There
is a calculation of the frequency dependence of the ORF in the Fig. 7 of CH14c,
and it shows that the frequency dependence is really small below 10 Hz, for a
0.4 rad angular separation. For larger separation like $\pi$ rad, our
calculation shows that the effect is changing the ORF by a fraction of $0.05$
.}. We also notice that  $\Gamma \left ( f \right )$ has the same dimension as
$T_{r/h}^{2}$, but can be different from it by one or two orders of magnitude
(see the next section for detailed calculations). 

Eq.~(\ref{eq:snr}) can be simplified for order-of-magnitude estimation:
\begin{eqnarray}
    \text{SNR} \simeq  \frac{3H_{0}^{2} \Omega _{\text{GW}}  \left ( f \right )}{4\pi ^{2}} \frac{\Gamma  \left ( f \right )  \sqrt{2T \Delta f}   }{ f^{3}S_{n}\left ( f \right )} ~, \label{eq:snrAPPROX}
\end{eqnarray}
where $\Delta f$ is the bandwidth. Therefore, the threshold $\Omega _{\text{GW}}$ that is possibly detectable would be
\begin{eqnarray}
    \Omega _{\text{GW}} \simeq \frac{4\pi ^{2}}{3H_{0}^{2}} \frac{f^{3}S_{n}\left ( f \right )}{\sqrt{2T \Delta f}}\frac{\text{SNR}}{\Gamma  \left ( f \right ) }  ~.\label{eq:omega}
\end{eqnarray}

By setting $\Delta f \simeq f$ and $\Gamma \simeq \eta T_{r/h}^{2}$, the above equation can be further simplified to:
\begin{eqnarray}
 &&\Omega _{\text{GW}} \simeq \frac{4\pi ^{2} }{3\sqrt{2} } \frac{\text{SNR}}{\eta} \frac{1}{H_{0}^{2}} \frac{f^{5/2}}{\sqrt{T }}\left ( \frac{\sqrt{S_{n}} }{T_{r/h}  }  \right ) ^{2} \nonumber \\ \simeq && 3.2 \times 10^{-7}\left ( \frac{\text{SNR}}{\eta}  \right ) \left ( \frac{H_{0} }{70\thinspace \text{km}\thinspace \text{s}^{-1}\thinspace \text{Mpc}^{-1}   }  \right ) ^{-2}  \left ( \frac{f}{1 \thinspace\text{Hz} }  \right )^{-3/2} \nonumber\\ \times&& \left ( \frac{T}{1 \thinspace\text{yr} }  \right ) ^{-1/2}   \left ( \frac{f^{2} \sqrt{S_{n}}   }{10^{-15}\thinspace \text{m}\thinspace \text{s}^{-2}\thinspace \text{Hz}^{-1/2}    }  \right ) ^{2}   \left ( \frac{T_{r/h} }{10^{5}\thinspace\text{m}  }  \right )^{-2}  ,
\end{eqnarray}
in which we have applied Eq.~(\ref{eq:patt}). The $f^{2} \sqrt{S_{n}}$ term
approximately corresponds to the sensitivity of the seismometer. Notice that
this equation generally underestimates the detectability by one or two orders
of magnitude, as we will discuss in Sec.~\ref{sec:appl}.

\section{\label{sec:ORF}Overlap reduction function}

In this section, we will calculate the ORFs using our new form of lunar
response solution. We begin with the ORFs for a network of two seismometers with
several special configurations. We next generalize the results to two
seismometers with arbitrary configurations, and end this section with the ORFs
for an array of multiple seismometers.

\subsection{\label{sec:ORF1}The ORF of two seismometers with special orientations}

We first calculate the ORFs for several special cases, for easier comparison with
the previous results.  
For the configuration considered in CH14c
[see Eq.~(\ref{eq:pattJH})], where
the seismometers are all vertical, we can calculate the ORF with
\begin{eqnarray}
    \Gamma^{\text{ver-ver}}  =&& \int \frac{\mathrm{d}^{2}\hat{n}  }{4\pi }  \int \frac{\mathrm{d} \psi }{2\pi } \sum_{A}F_{1}^{A,\text{CH14c}}\left ( \hat{n}  \right ) F_{2}^{A,\text{CH14c}}\left ( \hat{n}  \right ) \nonumber \\ = && T_{r}^{2} \int \frac{\mathrm{d}^{2}\hat{n}  }{4\pi } \int \frac{\mathrm{d} \psi }{2\pi } \sum_{A}\left ( \hat{e}_{r1} \cdot \mathbf{e }_{A}   \cdot \hat{e}_{r1} \right ) \left ( \hat{e}_{r2} \cdot \mathbf{e }_{A}   \cdot \hat{e}_{r2} \right ) \nonumber \\ = && \frac{2}{15}    T_{r}^{2}\left ( 1+3\cos 2\delta  \right )  ~. \label{eq:ORF_JH}
\end{eqnarray}
Here we have set $\hat{e}_{r1} = \left ( 0,0,1 \right ) $ and $\hat{e}_{r2} =
\left ( \sin \delta ,0,\cos \delta \right )  $, in which $\delta$ is the angle
subtended between the two seismometers by the great circle. Notice that in the
low-frequency approximation the ORF does not depend on frequency.  The maximum
of ORF occurs when $\delta = 0$ or $\pi$. Eq.~(\ref{eq:ORF_JH}) is consistent with
CH14a and CH14c. 

It is also interesting to consider two seismometers with horizontal accelerometers
(i.e., $\hat{e}_{\rm det} \cdot \hat{e}_{r} = 0$). The corresponding pattern functions
are  
 \begin{eqnarray}
     F^{\text{hor}}_{A} = 2T_{h} \left ( \hat{e}_{\rm det} \cdot \mathbf{e }_{A}   \cdot \hat{e}_{r} \right ), \quad A = + , \times ~.
 \end{eqnarray}
(i) If the two horizontal accelerometers are both vertical to the plane of the great circle between the two seismometers, the ORF is:
  \begin{eqnarray}
     \Gamma^{\text{hor1-hor1}}  = \pm  \frac{8}{5} T_{h}^{2} \cos \delta~,
 \end{eqnarray}
where $\pm$ represents two cases in which the 
the two accelerometers point in the same or the opposite directions, respectively. The maximum response occurs at $\delta = 0, \pi$.
(ii) If the two accelerometers are both tangent to a same great circle, the ORF is:
  \begin{eqnarray}
     \Gamma^{\text{hor2-hor2}}  = \pm  \frac{8}{5} T_{h}^{2} \cos 2\delta~,
 \end{eqnarray}
where $\pm$ represents the cases in which the two accelerometers point in the same or the opposite {\it arc} directions along the great circle. The maximum value occurs when $\delta = 0, \pi/2, \pi$.

We also consider other interesting cases in which the two seismometers have
different orientations. For example, if one accelerometer is vertical and
another is horizontal (relative to the surface of the moon), there are two kinds
of combinations depending on the direction of the horizontal detector. Using
the same notations `ver', `hor1', and `hor2' as before, we calculate the ORFs
and derive
 \begin{eqnarray}
     \Gamma^{\text{ver-hor1}}  =  0~,
 \end{eqnarray}
 \begin{eqnarray}
     \Gamma^{\text{ver-hor2}}  = \mp \frac{4}{5}  T_{r} T_{h} \sin 2 \delta~,
 \end{eqnarray}
where $\mp$ corresponds to the cases in which the horizontal accelerometer
aligned with the great circle is pointing in or opposite to the direction of
the vertical detector.  The maximum response occurs when $\delta = \pi/4,
3\pi/4$.

Now there is one kind of configuration that is still missing, 
in which the ORF is calculated as
 \begin{eqnarray}
     \Gamma^{\text{hor1-hor2}}  =  0~.
 \end{eqnarray}

\begin{figure}
\includegraphics[width=0.98\linewidth]{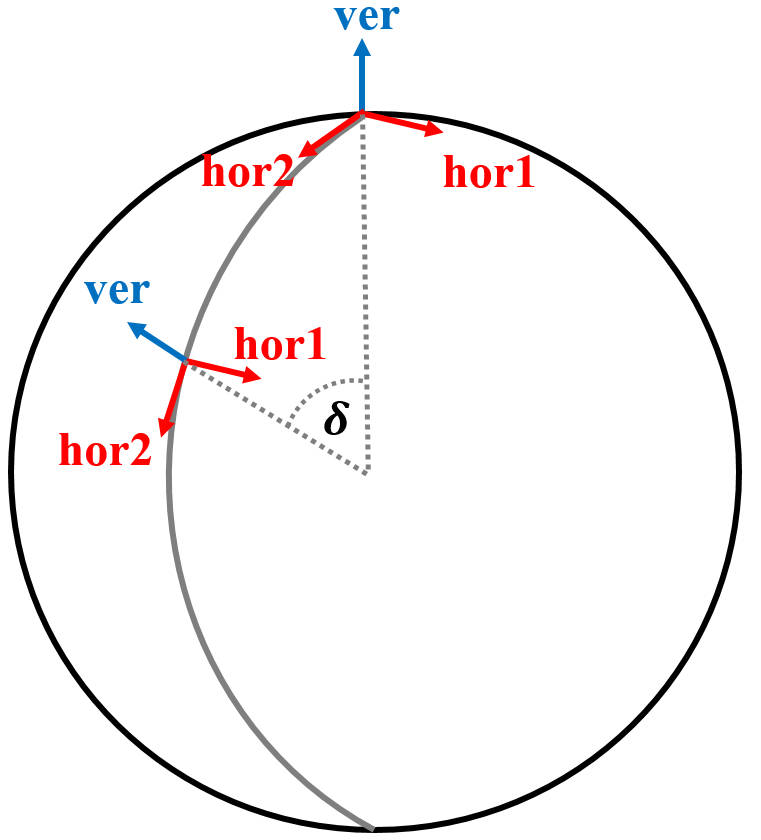}
\caption{\label{fig:config}Schematic diagram showing different configurations 
of the readout directions of the seismometers on the moon. The grey arc represents 
the great circle crossing two seismometers.}
\end{figure}

To better illustrate these configurations, we plot Fig.~\ref{fig:config} to show the meanings of ``ver'', ``hor1'' and ``hor2''.

By reviewing the above results, we find that in the configurations of
`ver-hor1' and `hor1-hor2', the ORFs are always zero. These results are due to
the cancellation of the response after the integration over the $4 \pi$ solid angle
and $2 \pi$ polarization angle.


\subsection{\label{sec:ORF2}The ORF of an array of seismometers with arbitrary 
orientations}

For two seismometers with arbitrary orientations, we can calculate the ORF according
to the linear dependence of the pattern function $F_{A}$ on the orientation
$\hat{e}_{\rm det }$ of the accelerometers inside the seismometers. For example,
consider two seismometers placed at the same position as described in
Sec.~\ref{sec:ORF1}. We can rename the base vectors of the orientations of the two
accelerometers as follows:
\begin{eqnarray}
    &&\hat{e}_{a1}  = \left ( 0,0,1 \right )  \nonumber \\ &&\hat{e}_{b1}  = \left ( 0,1,0 \right )  \nonumber \\ &&\hat{e}_{c1}  = \left ( 1,0,0 \right )  \nonumber \\ &&\hat{e}_{a2}  = \left ( \sin \delta ,0,\cos \delta \right )   \nonumber 
    \\ &&\hat{e}_{b2}  = \left ( 0,1,0 \right )  \nonumber 
    \\ &&\hat{e}_{c2}  = \left ( \cos \delta ,0,-\sin \delta \right )  ~, \label{eq:base}
\end{eqnarray}
where $a, b, c$ denote `ver', `hor1' and `hor2' respectively. 
Noticing that the orientation vectors can be decomposed as
\begin{eqnarray}
    &&\hat{e}_{\text{det}, i}  = A_{i} \hat{e}_{ai}+B_{i} \hat{e}_{bi}+C_{i} \hat{e}_{ci},\quad i=1,2 \\ &&A_{i}^{2} + B_{i}^{2} + C_{i}^{2} = 1 \nonumber ~,
\end{eqnarray}
we can write down the total ORF as
\begin{eqnarray}
    \Gamma = &&\frac{2}{15} \times [ A_{1}A_{2}  T_{r}^{2}\left ( 1+3\cos 2\delta  \right ) \nonumber \\ &&+ 12 B_{1}B_{2} T_{h}^{2} \cos \delta \nonumber \\ &&+ 12 C_{1}C_{2} T_{h}^{2} \cos 2\delta \nonumber \\ &&+ 6\left ( A_{2}C_{1}-A_{1}C_{2}     \right )   T_{r} T_{h} \sin 2 \delta ]~.
\end{eqnarray}
In particular, if we consider two seismometers installed nearby on the moon, which is the case of Chang'e seismometers \footnote{Chang'e 7 and 8 landers will both land near the lunar south pole.} and current design of LGWA \cite{2023JAP...133x4501V}, we can regard $\delta$ as a small quantity. 
The corresponding ORF can be calculated with
\begin{eqnarray}
    \Gamma^{\delta \sim  0} \simeq &&\frac{8}{15}  T_{r}^{2} \times \Bigg[ \sqrt{\left ( 1-B_{1}^{2}-C_{1}^{2}  \right ) \left ( 1-B_{2}^{2}-C_{2}^{2}  \right )}   \nonumber \\ &&+ 3 (B_{1}B_{2}+C_{1}C_{2}) \zeta ^{2}    \Bigg]~, \label{nearorf}
\end{eqnarray}
where $\zeta $ is a ratio defined by $T_{h}=\zeta T_{r}$.
In this specific situation 
the maximum value of ORF depends on $\zeta $. If $\zeta ^{2}> 1/3$, the maximum value is
$8\zeta ^{2}T_{r}^{2}/5$ when $B_{i} = C_{i} = 1/\sqrt{2},~i=1,2$.  Otherwise, if
$\zeta ^{2}< 1/3$, the maximum becomes $8T_{r}^{2}/15$ when $B_{i} = C_{i} =
0,~i=1,2$. For two seismometers $i$ and $j$ that are all horizontal (i.e., $A_{i}=A_{j}=0$), Eq.~(\ref{nearorf}) can be further simplified to
\begin{eqnarray}
    \Gamma^{\delta \sim  0,\rm hor}_{ij} &&= \frac{8}{5}  T_{h}^{2} (B_{i}B_{j}+C_{i}C_{j}) \nonumber \\ &&= \frac{8}{5}  T_{h}^{2} \hat{e}_{\text{det}, i} \cdot \hat{e}_{\text{det}, j} ~.\label{nearorf-hor}
\end{eqnarray}
We note that these ORFs [Eq.~(\ref{nearorf}) and Eq.~(\ref{nearorf-hor})] may
overestimate the SNR for a seismometer array in a small region, because two
seismometers placed nearby could have gained correlated environmental noise,
e.g., from the seismic wave induced by a nearby source of moonquake. This noise
will result in a worse sensitivity of the array to the GW background.  However,
this effect can be mitigated by adding another seismometer far from the
previous two, e.g., on the far side of the moon as the phase 2 of the LGWA
project proposes.

If there are more than two  seismometers, the SNR for such a detector array can be 
calculated with the following replacement in Eq.~(\ref{eq:snr}) \cite{2009GReGr..41.1667H}:
\begin{eqnarray}
    T\frac{\Gamma ^{2}\left ( f \right )  }{S_{n}^{2} \left ( f \right )  }  \to \sum_{i< j} T_{ij} \frac{\Gamma_{ij}  ^{2}\left ( f \right )  }{S_{n,i} \left ( f \right ) S_{n,j} \left ( f \right ) } ~,
\end{eqnarray}
where the summation is taken within different pairs ($i$-th and $j$-th seismometer) of seismometers, and $T_{ij}$ is the common operation time for each pair. If all the operational periods $T_{ij}$ are equal, and all the noise spectrum $S_{n,i}(f)$ are the same, we can calculate the SNR by defining an ORF of the multiple-seismometer array, 
which reads
\begin{eqnarray}
    \Gamma ^{\text{array}} = \sqrt{\sum_{i< j} \Gamma_{ij}  ^{2}} ~. \label{comb}
\end{eqnarray}
For example, for two seismometers in Chang'e 7 and 8 projects, each of them might be able to simultaneously measure the seismic responses from three orthogonal directions. Therefore, the ORF for Chang'e seismometers can be estimated as follow:
\begin{eqnarray}
    \Gamma ^{\text{Chang'e}} && \simeq \sqrt{\left ( \frac{8}{15}    T_{r}^{2} \right )^{2} + \left ( \frac{8}{5} T_{h}^{2} \right )^{2} + \left ( \frac{8}{5} T_{h}^{2} \right )^{2}  }  \nonumber \\ && = \frac{8}{15} \sqrt{T_{r}^{4} + 18 T_{h}^{4}}~, \label{ChangEorf}
\end{eqnarray}
where we have assumed that six readout directions just correspond to the base directions defined in Eq.~(\ref{eq:base}).

\begin{figure}
\includegraphics[width=0.98\linewidth]{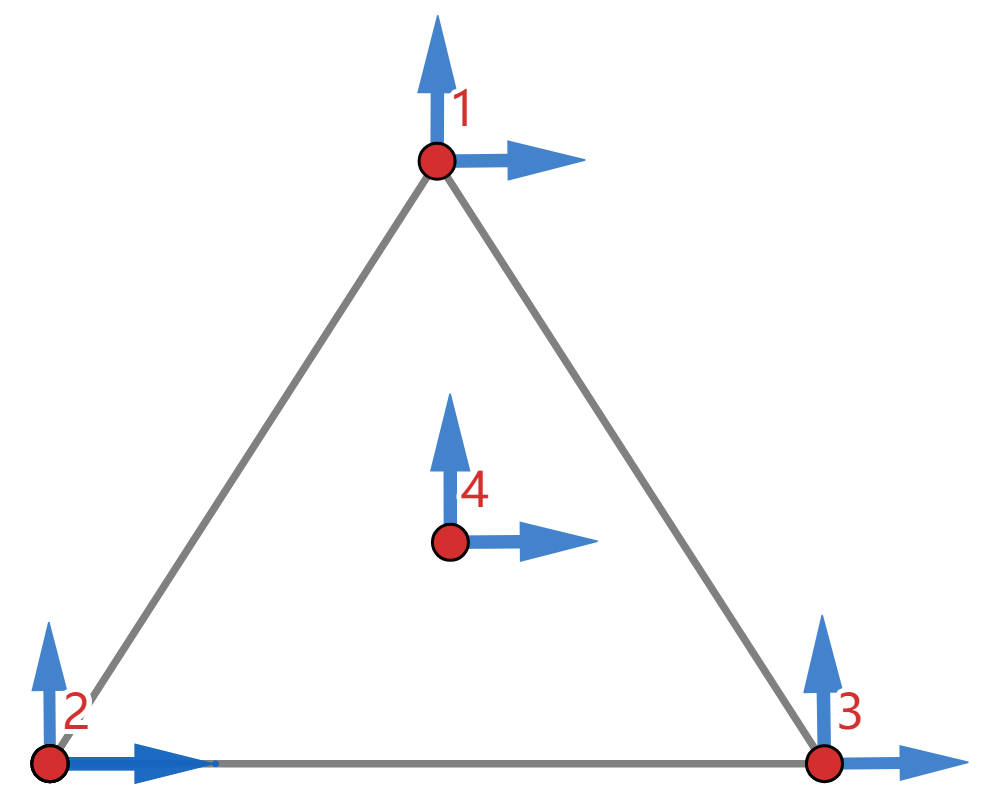}
\caption{\label{fig:0}Schematic diagram for the locations and orientations of LGWA seismometers. Short arrows remark the horizontal readout directions. The distances between these seismometers are around several kilometers, much smaller than the radius of the moon. }
\end{figure}

For LGWA, three seismometers are placed on the vertices of a small equilateral triangle (kilometer scale) and one seismometer is placed at the center of this triangle (all seismometers have two orthogonal horizontal readouts). We plot Fig.~\ref{fig:0} as a sketch for their locations and directions (notice that we have specified the readout directions of each seismometer for simplicity, which might not be the real case). In this case, we first calculate the ORF between each pairs:
\begin{eqnarray}
    \Gamma_{12} = \Gamma_{13} = \Gamma_{14} = \Gamma_{23} = \Gamma_{24} = \Gamma_{34} = \frac{8}{5}\sqrt{2} T_{h}^{2} ~.
\end{eqnarray}
The $8T_{h}^{2}/5$ term in the above equation comes from $x-x$ or $y-y$ correlation in each pair of the seismometers, which can be easily calculated using Eq.~(\ref{nearorf-hor}), and the combination of these correlations leads to an extra factor of $\sqrt{2}$, according to Eq.~(\ref{comb}). 
As a result, the ORF for LGWA seismometer array becomes
\begin{eqnarray}
    \Gamma ^{\text{LGWA}} \simeq \sqrt{C_{4}^{2}}  \times \frac{8}{5} \sqrt{2 }  T_{h}^{2} = \frac{16\sqrt{3} }{5} T_{h}^{2}   ~. \label{LGWAorf}
\end{eqnarray}

\section{\label{sec:appl}Constraining the SGWB by lunar seismometers}

\begin{figure}
\includegraphics[width=0.98\linewidth]{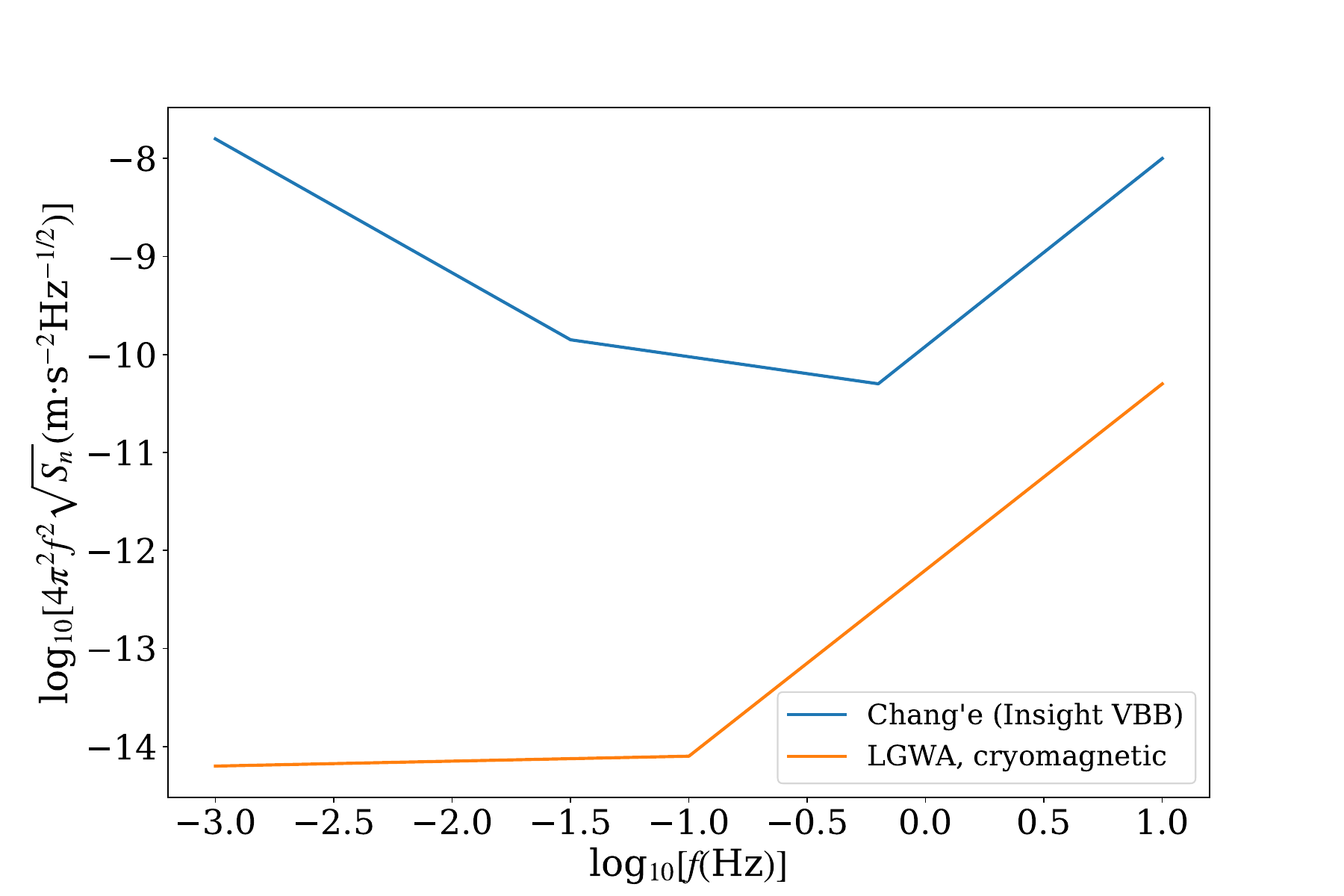}
\caption{\label{fig:1}The sensitivities of the seismometers 
in two future projects, namely Chang'e (approximated by the Insight VBB) and LGWA cryomagnetic design.}
\end{figure}

In this section, we will evaluate the viability of using lunar seismometers to
constrain the SGWB. We mainly consider two different projects: one is the
Chang'e project from China and  the other is the LGWA project from Europe. We
approximate the sensitivity of the seismometers of Chang'e 7 and 8 by the
sensitivity of the Insight's Very Broad Band (VBB) seismometer on Mars
\cite{2019SSRv..215...12L}, and for LGWA we choose the cryomagnetic design
\cite{2021ApJ...910....1H}. The sensitivity curves of these seismometers are plotted in
Fig.~\ref{fig:1}.

\begin{figure}
\includegraphics[width=0.98\linewidth]{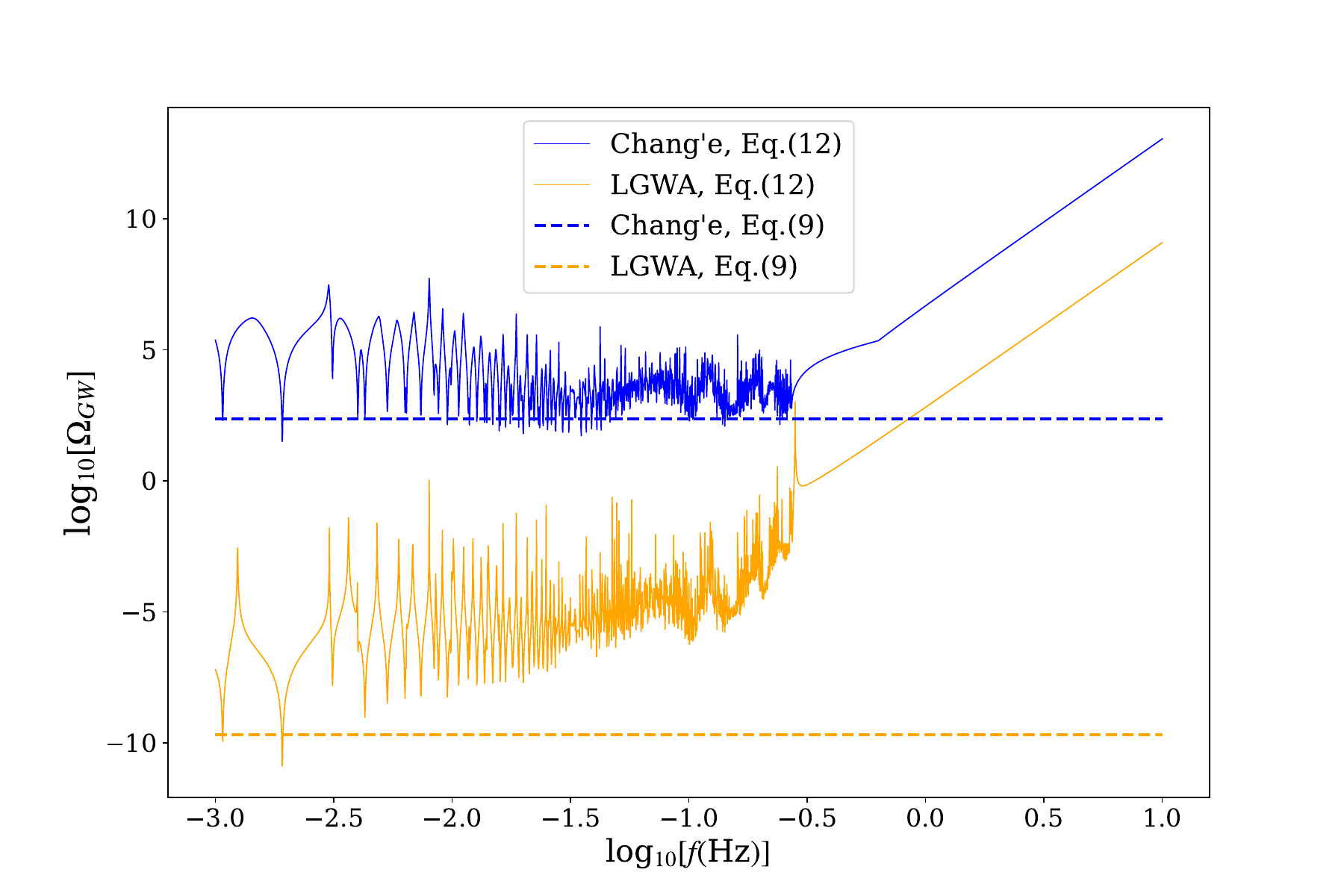}
\caption{\label{fig:2}Constraints on the SGWB by Chang'e (blue) and LGWA (orange). 
The results from a
full integration of Eq.~(\ref{eq:snr}) are shown as the dashed horizontal lines. 
The approximate results calculated by Eq.~(\ref{eq:omega}) are plotted in solid lines.}
\end{figure}

Next, we calculate the constraint on the energy spectral density of the SGWB by
Chang’e and LGWA. According to Eqs.~(\ref{ChangEorf}) and (\ref{LGWAorf}) and
the related discussions in Sec.~\ref{sec:ORF2}, we set
$\Gamma^{\text{Chang'e}}=8 \sqrt{T_{r}^{4} + 18 T_{h}^{4}}/15$ and
$\Gamma^{\text{LGWA}}=16\sqrt{3} T_{h}^{2}/5$.  Then the detection thresholds are
derived in two different ways. (i) First, we use the approximation,
Eq.~(\ref{eq:omega}), and set $\Delta f = f$, $\text{SNR}=3$, and $T = 1
\text{yr}$.  The results are shown in Fig.~\ref{fig:2} as the solid lines.
(ii) Second, we perform a more accurate integration using Eq.~(\ref{eq:snr}), where
we have assumed a flat spectrum for $\Omega_{\text{GW}}$ [i.e., setting $\Omega_{\text{GW}}(f)$ = $\Omega_{\text{GW}}$], and set
$\text{SNR}=3$ and $T = 1 \text{yr}$.  The integration is performed in the
frequency domain of $10^{-3} - 10~\text{Hz}$.  The corresponding thresholds are
shown in Fig.~\ref{fig:2} as the horizontal dashed lines.  The values
corresponding to these two dashed lines are
$\Omega_{\text{GW}}^{\text{Chang'e}} = 2.4 \times 10^{2}$ and
$\Omega_{\text{GW}}^{\text{LGWA}} = 2.0 \times 10^{-10}$, respectively. It is worth mentioning that our
approach can be easily extended to the cases with different shapes of SGWB energy spectrum, by changing the constant $\Omega_{\text{GW}}$ to a certain kind of function $\Omega_{\text{GW}}(f)$ in Eq.~(\ref{eq:snr}). In general, the sensitivity of LGWA might enable people to distinguish the differences between different SGWB models. 


\begin{figure}
\includegraphics[width=0.98\linewidth]{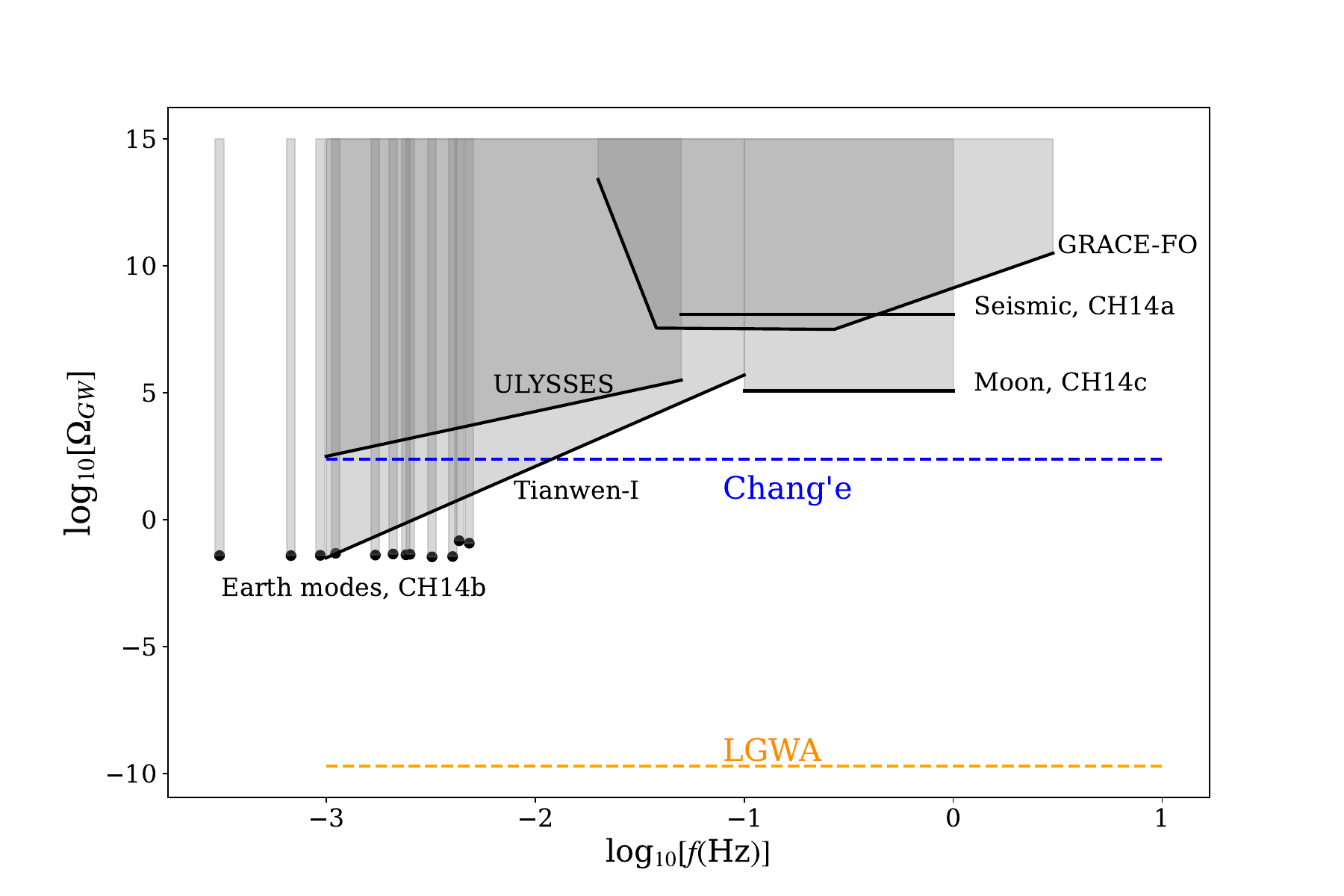}
\caption{\label{fig:3}Constraints on the SGWB 
given by Chang'e and LGWA (this work), 
as well as by ULYSSES \cite{1995A&A...296...13B}, Earth's normal modes (CH14a), terrestrial seismic motion (CH14b), lunar seismic motion (CH14c), GRACE-FO \cite{2020PhRvD.101j2004R}, and Tianwen-I \cite{2024arXiv240206096B}. Following the style in \cite{2020PhRvD.101j2004R}, gray shaded area indicates 
the excluded region.}
\end{figure}

For comparison, we also plot in Fig.~\ref{fig:3} the constraints on the SGWB in
the mid-frequency band given by previous works.  We find that in the
frequency band around $0.1 \text{Hz}$, the constraint given by Chang'e will be
better than the previous ones by about two orders of magnitude.  The constraint
by LGWA will greatly exceed any current constraints in the frequency band of
$10^{-3}-1$ Hz, even though the threshold $\Omega_{\text{GW}}$ derived in this
work is slightly worse than the previous prediction \cite{2021ApJ...910....1H}
because of the more up-to-date response functions used here. Again, we would like to mention that because two seismometers placed nearby could have gained
correlated environmental noise, our results might overestimate the detectability of SGWB.

\section{\label{sec:con}Summary and discussions}

Motivated by the recent improvements in the calculation of the
response of the moon to GWs \cite{PhysRevD.109.064092,2024arXiv240316550B,2024arXiv240305118B}, we revisited in this paper the detectability of the SGWB by lunar
seismometers. Besides applying the updated response functions,
we paid special attention to the effect imposed by the orientations of the
seismometers, which are not necessarily pointing in a direction vertical to the
moon surface due to different deployment mechanisms.  To evaluate such an
effect, we derived the pattern functions of a single seismometer for the two GW
polarizations (Sec.~\ref{sec:resptheo}). We note that these pattern functions will be
useful for future studies of the localization of GW sources by lunar
seismometer arrays.

Using the pattern functions, we also constructed the ORFs for a network of two seismometers (Sec.~\ref{sec:ORF1}) as well as an
array of an arbitrary number of seismometers (Sec.~\ref{sec:ORF2}).  We applied our
ORFs to two future projects, namely, Chang'e and LGWA (Sec.~\ref{sec:appl}). We
found that in the frequency band around $0.1$ Hz, the threshold SGWB detectable
by Chang'e is two orders of magnitude better than the limit given previously by
other missions. The sensitivity of LGWA would be even better, reaching  a level
as low as $\Omega_{\text{GW}}^{\text{LGWA}} \sim 2.0 \times 10^{-10}$
throughout the mid-frequency band of $10^{-3}-10$ Hz (Fig.~\ref{fig:3}).

Finally, we point out several caveats in this work which deserve future
investigation.  First, we have omitted the correlation between the noise of the
seismometers, which will be crucial for the seismometers placed at close
locations.  This correlation may undermine the
ability of a seismometer array in constraining the SGWB.  Second, as
we have mentioned in Yan24, we ignored the complex structure at the surface of the
moon, which might lead to strong scattering of the seismic waves \cite{2012JGRE..117.6003B,2022JGRE..12707222Z}.  This structure might influence the lunar response functions \cite{2024arXiv240305118B}, and should be examined by the
real data from future lunar seismometers. Third, a recent theoretical study of the
interaction between GW and elastic body hints that the current model of the
lunar response may be incomplete \cite{2024arXiv240316550B}. The difference
between the results deserves further investigation. 


\begin{acknowledgments}
This work is supported by the National Key Research and Development Program of
China Grants No. 2021YFC2203002, No. 2023YFC2205800, the Beijing Natural Science Foundation Grant No. 1242018, the Fundamental Research Funds for the Central Universities Grant No. 310432103, and the National Natural Science Foundation of China (NSFC) Grants No. 11991053, No. 12073005, No. 12021003, No. 42325406. The authors would like to thank Jan Harms for helpful discussions.
\end{acknowledgments}

\appendix

\section{\label{app:GWproj}Proving that Eq.~(\ref{eq:resp}) applies to both polarizations} 

We first consider a new polarization state ($h_{+} - h_{\times}$) of GW that is different from 
Eq.~(\ref{GWtensor}). It can be written as
\begin{equation}
	\mathbf{h}_{\rm new} =  \Re \left \{ h_{0} \epsilon _{ij,\rm new} e^{i\left ( \omega _{g}t - \vec{k}_{g}\cdot \vec{r}     \right ) }    \right \},
\end{equation}
where
\begin{eqnarray}
\epsilon _{ij, \rm new} && = \begin{bmatrix}
 1 & -1 & 0\\
 -1 & -1 & 0\\
 0 & 0 & 0
	\end{bmatrix}.
\end{eqnarray}

To prove that this type of polarization also satisfies 
Eq.~(\ref{eq:resp}), the key step is to prove
that it satisfies Eq.~(25) in Yan24. The latter proof requires us
to calculate the function $f^{m}_{\rm new} $, which depends on the wave vector
direction and polarization state (i.e., angles $e,\lambda$ and $\nu$ in Ma19).
Notice that in Ma19 and Yan24 the function $f^{m} $ was calculated only for a
specific polarization state. 

Considering the definitions of angles $e,\lambda$
and $\nu$, we find that $f^{m}_{\rm new}$ can be obtained by replacing these
angles with $(e+\pi),\lambda$, and $(\pi-\nu)$ respectively, which results in
$\vec{l} \to  \vec{l}, \vec{m} \to  -\vec{m}$, and $\hat{e} _{k}  \to  -\hat{e}
_{k} $.
As a result, we get:
\begin{eqnarray}
    f^{m}_{\rm new}  = && f^{m} \left ( e= \pi,\lambda = 0, \nu =\pi \right )  \nonumber \\ = && 4\sqrt{\frac{\pi }{15} } \times \left (- \delta_{m, 2}+  \delta_{m,-2} \right ).
\end{eqnarray}

Now we can combine the previously derived functions. On one hand,
using the definitions of the three base vectors, we can derive
\begin{eqnarray}
    \hat{e}_{r}\cdot \epsilon_{\rm new} \cdot \hat{e}_{r}&& = \sin^{2} \theta \left (- \sin 2\varphi + \cos 2\varphi\right ), \nonumber \\
    \hat{e}_{\theta }\cdot \epsilon_{\rm new} \cdot \hat{e}_{r}&& = \sin \theta \cos \theta \left ( -\sin 2\varphi + \cos 2\varphi\right ), \nonumber \\
    \hat{e}_{\varphi  }\cdot \epsilon_{\rm new} \cdot \hat{e}_{r} &&= \sin \theta  \left (- \sin 2\varphi - \cos 2\varphi\right ) ~.
\end{eqnarray}
On the other hand, 
considering the definition of the real spherical harmonics 
and $f^{m}_{\rm new}$, we have:
\begin{eqnarray}
    \sum_{m} \mathcal{Y}_{2m} \left ( \theta ,\varphi  \right )f^{m}_{\rm new} &&= \sin^{2}\theta\left ( -\sin 2\varphi + \cos 2\varphi  \right ) ,\nonumber \\
    \sum_{m} \partial_{\theta }  \mathcal{Y}_{2m} \left ( \theta ,\varphi  \right )f^{m}_{\rm new} &&= \sin2\theta\left ( -\sin 2\varphi + \cos 2\varphi  \right ) ,\nonumber \\
    \sum_{m} \frac{\partial _{\varphi  }  \mathcal{Y}_{2m} \left ( \theta ,\varphi  \right )}{\sin \theta }f^{m}_{\rm new} &&= 2\sin\theta\left ( -\sin 2\varphi - \cos 2\varphi  \right )  ~.
\end{eqnarray}

Therefore, we find
\begin{eqnarray}
    \frac{1}{2} \hat{e}_{r} \cdot   \mathbf{h} _{\rm new}  \cdot \hat{e}_{r} && = \frac{h_{0}  \cos \left ( \omega _{g}t  \right )}{2}    \sum_{m} \mathcal{Y}_{2m} \left ( \theta ,\varphi  \right )f^{m}_{\rm new}, \nonumber \\
    \frac{1}{2} \hat{e}_{\theta } \cdot   \mathbf{h}_{\rm new}   \cdot \hat{e}_{r} &&= \frac{h_{0}  \cos \left ( \omega _{g}t  \right )}{4}   \sum_{m} \partial_{\theta }  \mathcal{Y}_{2m} \left ( \theta ,\varphi  \right )f^{m}_{\rm new}, \nonumber \\
    \frac{1}{2} \hat{e}_{\varphi } \cdot   \mathbf{h}_{\rm new}   \cdot \hat{e}_{r} &&= \frac{h_{0}  \cos \left ( \omega _{g}t  \right )}{4}    \sum_{m} \frac{\partial _{\varphi  }  \mathcal{Y}_{2m} \left ( \theta ,\varphi  \right )}{\sin \theta } f^{m}_{\rm new}  . \label{eq:proj}
\end{eqnarray}
This concludes the proof that Eq.~(25) in Yan24 is still 
satisfied when we consider a different polarization state.



\bibliography{main}

\end{document}